\documentclass[acmsmall,screen,nonacm]{acmart}

\usepackage{multirow}
\usepackage{makecell, graphicx}

\AtBeginDocument{%
  }

\newcommand{\tool}[0]{\textsc{BScout}}

\newcommand{\binkit}[0]{BinKit-BCSA}

\title{Recall-Oriented Precise Binary-to-Source Function Matching}
\author{Junlin Li}
\email{lijunlin@ruc.edu.cn}
\affiliation{%
  \institution{Renmin University of China}
  \city{Beijing}
  \country{China}}

\author{Jianjun Huang}
\email{hjj@ruc.edu.cn}
\affiliation{%
  \institution{Renmin University of China}
  \city{Beijing}
  \country{China}}

\author{Wei You}
\email{youwei@ruc.edu.cn}
\affiliation{%
  \institution{Renmin University of China}
  \city{Beijing}
  \country{China}}

\author{Bin Liang}
\email{liangb@ruc.edu.cn}
\affiliation{%
  \institution{Renmin University of China}
  \city{Beijing}
  \country{China}}

\renewcommand{\shortauthors}{Li et al.}

\setcopyright{none}
\acmDOI{}
\acmISBN{}
\hypersetup{keeppdfinfo}
\makeatletter
\@ifpackageloaded{hyperxmp}{\let\hyxmp@parse@acmart\relax}{}
\makeatother

\begin{document}

\begin{abstract}

Binary-to-source function matching supports software composition analysis by identifying reused third-party code, but existing embedding-based approaches restrict subsequent refinement to initially retrieved candidates. When the correct source function is absent from this set, selecting or filtering the available candidates cannot recover the match. Inlining further complicates the retrieval by combining multiple source functions into one binary function. In this paper, we propose \tool{} to improve matching precision and uncover additional matches. \tool{} uses the positional information of existing matches to supplement candidates for unresolved binary functions, extending the search beyond the initial candidate set, and adopts composition-aware training to address the inlined composition matching. Large language models are adopted to select and review matches, while positional information also helps filter inconsistent results. Experiments show that \tool{} is effective in binary-to-source function matching. \tool{} achieves overall higher precision, recall and F1 scores than BinaryAI and SBridge with acceptable analysis costs. 
Particularly, on large stripped binaries, \tool{} outperforms the competitors with a precision increment of 32\% and approximately 2.9 times the recall.  
\end{abstract}




\maketitle

\section{Introduction}\label{sec-introduction}

Modern software extensively reuses open-source projects and third-party code~\cite{choi2025tiver}. When the software source code is unavailable, matching the binary to reference source code can provide evidence for software composition analysis (SCA)~\cite{jiang2024binaryai}, license compliance checking~\cite{duan2017osspolice}, and vulnerability analysis~\cite{yang2026sbridge,li2025lares}. %
Binary-to-source function matching~\cite{yu2020codecmr} is the fundamental technique for conducting the analysis, which addresses the problem of retrieving the corresponding source function from a reference corpus of source functions for each binary query function in a given binary file.
There have been a lot of binary-to-source function matching approaches. B2SFinder~\cite{feng2019b2sfinder} utilizes the syntactic features, e.g., string literals, which can be redundant across a large number of third-party libraries (TPLs) and harm the matching precision~\cite{jiang2024binaryai}. Beyond the syntactic features, SBridge~\cite{yang2026sbridge} leverages structural and semantic features to achieve more accurate matching, but the time cost on a large corpus is unacceptable. Embedding-based matching is a solution to address both the efficiency and effectiveness pits of the above feature-engineering approaches. It leverages the ability of deep learning to automatically learn the features from the code and improve the matching efficiency based on vector similarity measurement~\cite{yu2020codecmr,jiang2024binaryai}. BinaryAI~\cite{jiang2024binaryai} is the state-of-the-art (SOTA) method in this area. However, existing embedding-based methods such as BinaryAI~\cite{jiang2024binaryai} and CodeCMR~\cite{yu2020codecmr} overlook a common phenomenon in C/C++ programs, i.e., the inlining mechanism combines more than one source function into one binary function~\cite{abusabha2026inlining}, preventing them precisely and completely pinpointing the correct binary-source function matches. 
Furthermore, they restrict their final results within the scope of the embedding-based matching results, losing the potential of finding more correct matches.

In this paper, we present \tool{}, a binary-to-source function-level matching approach that guarantees high precision and attempts to uncover more results beyond the embedding-based matching. \tool{} adopts embedding-based matching to enforce acceptable efficiency for large-scale matching tasks. \tool{} decompiles binary functions and parses source functions, converting each function into an embedding vector using a trained embedding model, measuring the binary-source function embedding similarity and emitting a set of candidates for each binary function. Then \tool{} adopts a large language model (LLM) and the positional information of the functions to iteratively refine the candidates, including filtering out improper matches and digging out new candidates for unmatched binary functions. At the end, \tool{} outputs a 1-to-1 mapping showing the binary-source function-level matching.

\tool{} addresses the two shortcomings of existing similar embedding-based approaches from the following two aspects.
First, we propose a composition-aware training, enforcing the embedding model to recognize each inlined composition in the binary function by weighting all the compositions and the major contributors. In addition, we design a chunk-based full-function embedding approach on top of the transformer-based model, to support and accelerate long function handling and training.
Second, we leverage the positional information not only to filter the matches but also to supplement additional candidates for unresolved binary functions. Such functions may have their matches outside the given candidates or rejected in the LLM selection steps. It is the essential effort in this study to expand matching results beyond the embedding-based matching results.

We evaluated \tool{} on a public dataset~\cite{yang2026sbridge} %
and a manually collected dataset with large software as binary and popular TPLs as source corpus. 
\tool{} has shown great improvement for retrieving inlined compositions for multi-source functions and uncovers 10,032 correct matches beyond the initial candidates.
On the datasets, it achieves the best comprehensive performance, high precision in all cases and enhanced recall over the embedding-based matching results.
It also outperforms the SOTA competitors, BinaryAI~\cite{jiang2024binaryai} and SBridge~\cite{yang2026sbridge} on each dataset with significant precision improvement (at least $\uparrow$32\%) and observable higher recall (about 2.9 times on large stripped binaries). 
Even if a small, locally deployed LLM is adopted, \tool{} shows obviously better performance. 
The efficiency and cost are also acceptable, making a large-scale matching affordable. 

Our paper makes the following contributions.

\begin{enumerate}
    \item We propose a binary-to-source function matching method, \tool{}. It leverages LLMs and the positional information to refine the embedding-based matching results, ensuring high precision and no recall degradation. 
    \item We design a composition-aware training for a transformer-based model to cope with the inlining mechanism that can make the binary function differ significantly from the corresponding source functions. More importantly, we present a method to supplement candidates and boost recall after the embedding-based matching. 
    \item We evaluate \tool{} on real-world datasets. It always achieves sufficiently high precision and best overall recall and comprehensive matching performance. \tool{} outperforms SOTA approaches with acceptable time and monetary cost.
\end{enumerate}

\section{Motivation}\label{sec-background-motivation}

In this section, we present two examples to motivate our techniques.\space

\subsection{Function Retrieval under Multi-Source Composition}\label{sec-background-multisource}

Inlining is common in C/C++ programs, which combines multiple source functions into a single binary function~\cite{jia2022o2nmatcher,jia2023inlining,jia2024crossinlining,yang2026sbridge}. A recent study~\cite{abusabha2026inlining} reports a mean function-inlining ratio of 32.77\% under the O2 optimization in its benchmark. We have also observed a comparable inlining ratio in the collected complicated software (see \S\ref{sec:datasets}). \space

\begin{figure}
\centering
\includegraphics[width=1\linewidth,height=\textheight,keepaspectratio,clip,trim=5 25 5 40,alt={The left side shows three source functions from Darktable: an entry function, a guide-drawing function, and an inline line-drawing helper. The entry function calls the guide-drawing function, which calls the helper four times. Their normalized contributions are 0.1177, 0.3604, and 0.5219, respectively; the latter two are major source functions. The right side shows the corresponding Ghidra-decompiled binary function FUN\_003ea5c0. Colored highlights connect operations in the source functions to the coordinate calculations and repeated Cairo calls in the decompiled code after inlining.}]{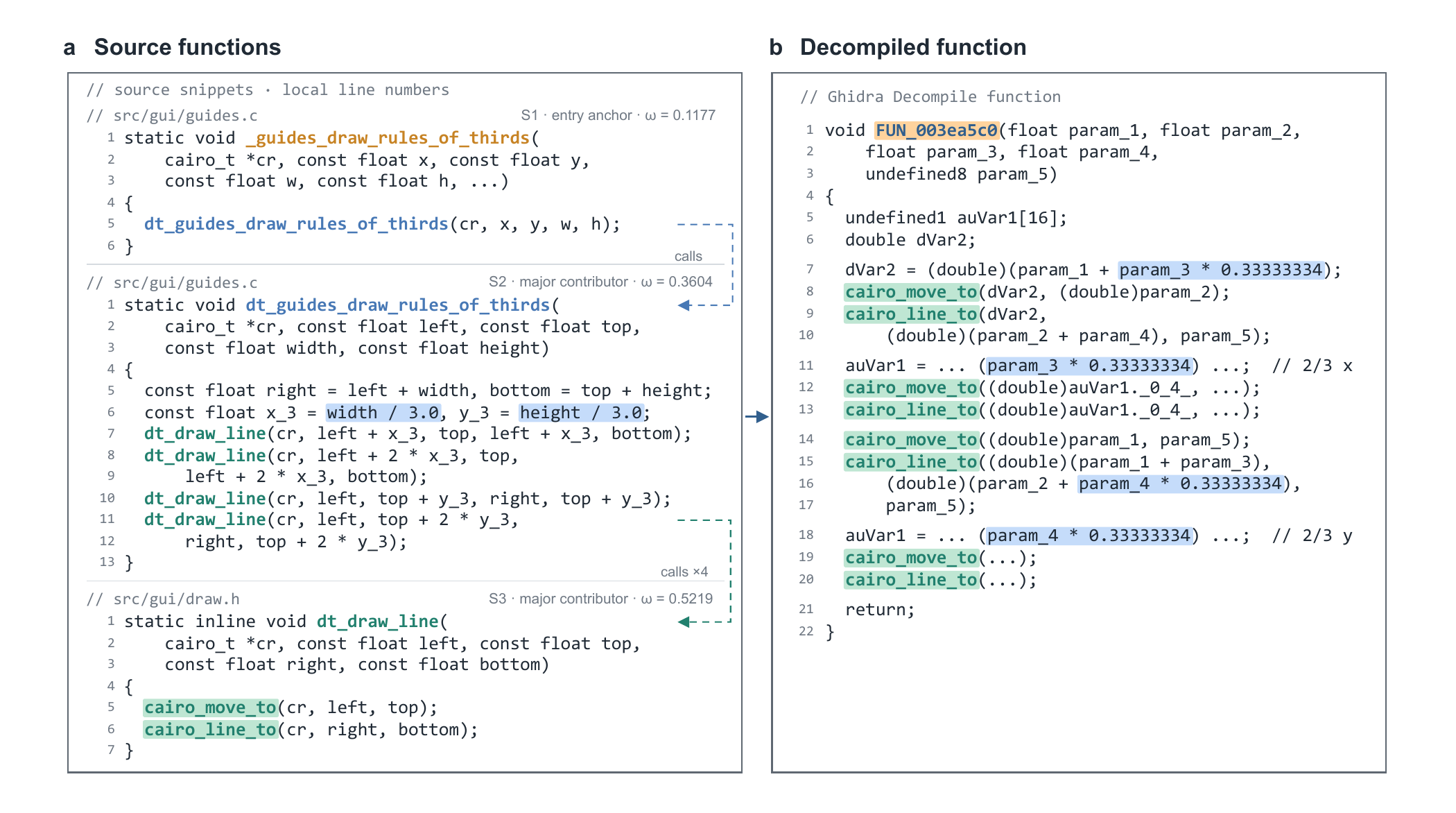}
\caption{Multi-source function composition in a real-world example, comprising three functions.}\label{fig-background-darktable-multisource}
\Description{The left side shows three source functions from Darktable: an entry function, a guide-drawing function, and an inline line-drawing helper. The entry function calls the guide-drawing function, which calls the helper four times. Their normalized contributions are 0.1177, 0.3604, and 0.5219, respectively; the latter two are major source functions. The right side shows the corresponding Ghidra-decompiled binary function FUN\_003ea5c0. Colored highlights connect operations in the source functions to the coordinate calculations and repeated Cairo calls in the decompiled code after inlining.}
\end{figure}

A concrete example from a stripped binary of real-world software, i.e., darktable, is shown in Figure~\ref{fig-background-darktable-multisource}. The decompiled function by Ghidra~\cite{ghidra} is displayed along with the involved source functions. The binary function, decompiled by Ghidra~\cite{ghidra}, is composited with three source functions, one entry and two implementations. The entry function calls the second one, which calls the third one four times. Inlining all the callees results in a binary function much bigger than each of the functions. In addition, the calls in the entry and the second functions are eliminated, while the calls in the third function appear repeatedly in the binary function. \space 

Embedding-based function-level matching encounters a problem. The binary function, though without any complicated control-flow structures, is full of noise information compared to the three source functions: nothing to be similar with the entry, some similar operations but completely different calls with the second, and same calls as in the third but multiple copies and extra computation operations. Quite dissimilar presentations make the embedding-based retrieval difficult. Moreover, training a model with the supervision that the binary function is only similar to the entry or either of the other two, as many existing works have done~\cite{jiang2024binaryai,yu2020codecmr}, may force the model to ignore the relation between the binary and the other two functions.\space

We propose the composition-aware training, letting the model learn the relation between the binary function and each of its compositions. The training also takes the major compositions that form the majority of the binary function into consideration. As a result, at the embedding-based matching phase, both the second and the third functions are ranked within top 5 among one million source functions.\space

\begin{figure}
\centering
\includegraphics[width=1\linewidth,height=\textheight,keepaspectratio,clip,trim=5 32 5 20,alt={The three code columns show the initially selected false candidate LMMS IPAddress::compare on the left, the decompiled binary function FUN\_00397c70 in the middle, and the correct source function Transmission tr\_address::compare on the right. The false candidate is ranked tenth with a similarity of 0.722. The correct source function is ranked twentieth with a similarity of 0.693 and is absent from the initial candidate set. Highlighted code contrasts IPv4-mapped IPv6 address conversion with type comparison and four- or sixteen-byte memory comparisons. The lower workflow shows BScout filtering out the false match because its anchor is matched to a different TPL from the surrounding anchors in the Transmission net.cc interval. The binary function is then left unresolved between net.cc anchors. Using this positional information, BScout supplements candidates from net.cc, and the LLM selects tr\_address::compare as the correct match.}]{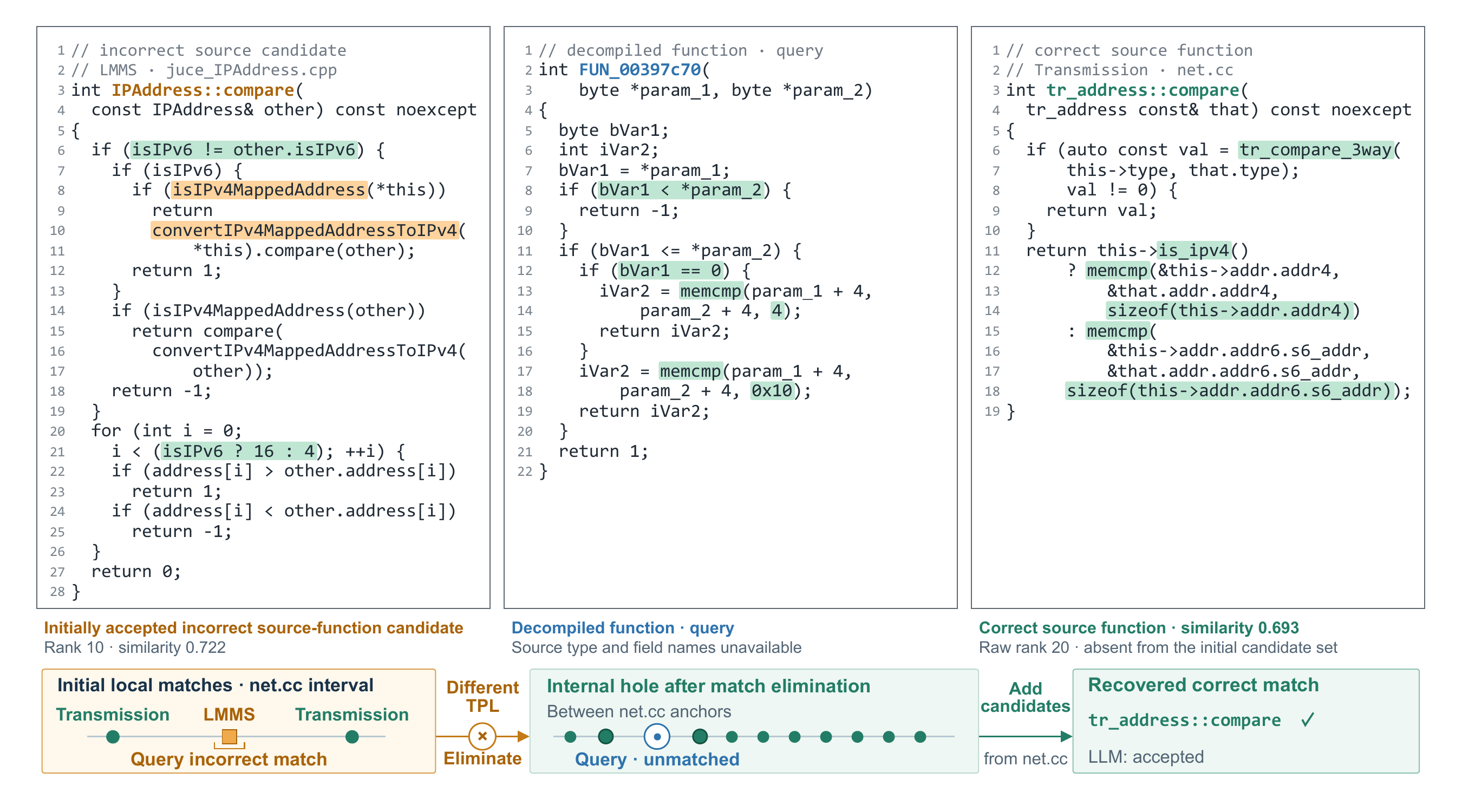}
\caption{A false candidate (left) is selected by the LLM for the binary function (middle) but later filtered out. Supplementing the correct candidate (right) emits a finally precise match.}\label{fig-background-transmission-recovery}
\Description{The three code columns show the initially selected false candidate LMMS IPAddress::compare on the left, the decompiled binary function FUN\_00397c70 in the middle, and the correct source function Transmission tr\_address::compare on the right. The false candidate is ranked tenth with a similarity of 0.722. The correct source function is ranked twentieth with a similarity of 0.693 and is absent from the initial candidate set. Highlighted code contrasts IPv4-mapped IPv6 address conversion with type comparison and four- or sixteen-byte memory comparisons. The lower workflow shows BScout filtering out the false match because its anchor is matched to a different TPL from the surrounding anchors in the Transmission net.cc interval. The binary function is then left unresolved between net.cc anchors. Using this positional information, BScout supplements candidates from net.cc, and the LLM selects tr\_address::compare as the correct match.}
\end{figure}

\subsection{Reliable Matching with Incomplete Candidate Sets}\label{sec-background-structural-completion}

Embedding-based matching typically emits top-$N$ candidates for the queried binary functions, and a similarity threshold is usually applied for precision consideration. In practice, no matter how the embedding model is trained, many binary functions can have their correct matches outside the candidates, thus hindering their discovery. The proposed \tool{} leverages LLMs to more precisely identify the correct match from the given candidates of a binary function, but the complication of the code and the limited capability of LLMs may miss the match even if it is there, or select a false one that can be easily filtered out later. \space

Figure~\ref{fig-background-transmission-recovery} shows an example. The embedding-based matching leaves 10 candidates for the binary function (in the middle), which indeed belongs to a TPL \textit{Transmission}. Among the 10, the LLM reports the 10th, i.e., the one at the left in Figure~\ref{fig-background-transmission-recovery}, is the correct match. But the real match should be the function at the right, which is ranked 20th, outside the candidates. \space 

The false match and the binary function differ significantly in the called functions, but we cannot blame the LLM for picking the false one from the candidates instead of rejecting all. Both the false one (left) and the correct one (right) implement the comparison semantics. That means, the left function has similar semantics with the binary function, regarding their design objectives.\space

\tool{} applies a simple filter after the LLM decision and discards the false match because the source function belongs to a different TPL (i.e., LMMS). Existing embedding-based similarity retrievals~\cite{jiang2024binaryai} focus only on the candidates, and even if some subsequent analysis is adopted, the discarded ones will never be re-considered.\space

We, in this paper, propose a strategy of supplementing the candidates for the unmatched binary functions based on their positional information. For the example, we can confidently pick some candidates, even though we do not know which one should be correct at this point, from a specific source file in Transmission, because the nearby matches are all associated to the file in the TPL. After supplementing the candidates (including the right function) for the query, we have the LLM make a decision again and it selects the correct one.\space

\section{Approach}\label{sec-method}

\subsection{Overview}

\begin{figure}[!t]
\centering
\includegraphics[width=1\linewidth,height=\textheight,keepaspectratio,clip,trim=0 18 0 18,alt={The BScout workflow consists of preprocessing and two matching stages. During preprocessing, tree-sitter extracts source functions from source code, and Ghidra decompiles the target binary into functions. Stage 1 maps source and decompiled functions to full-function embeddings, computes their similarity, and retrieves top-N source candidates for each binary function. Stage 2 combines position-driven filtering and supplementation with LLM-assisted selection and review to produce the matching results.}]{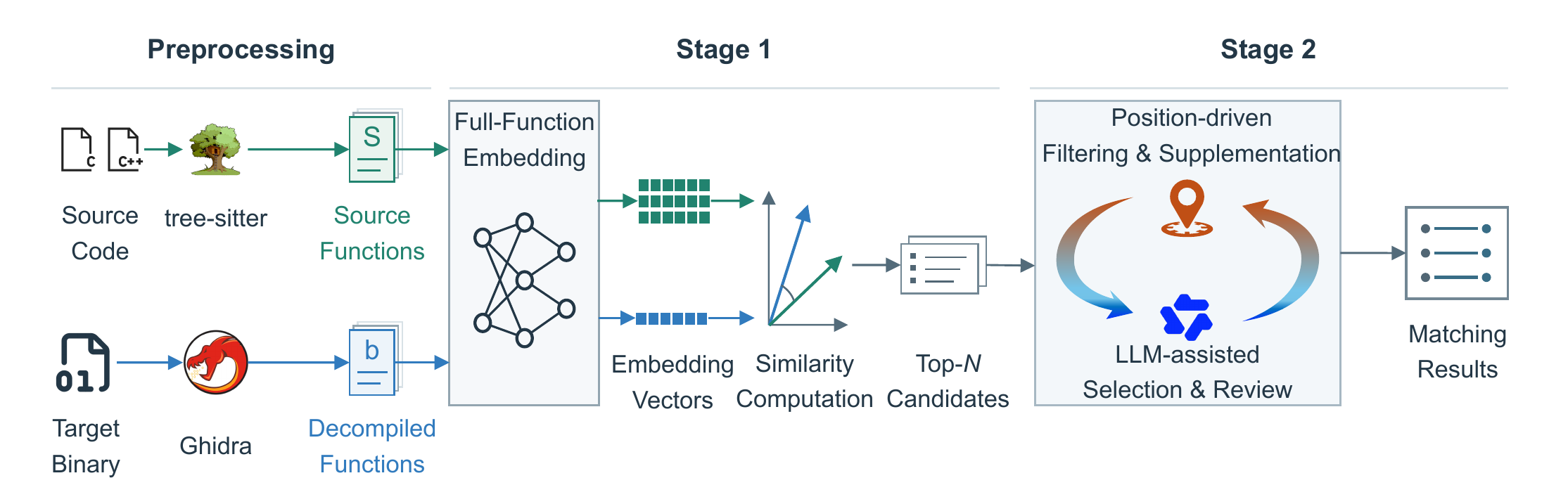}
\caption{Overview of \tool{}.}\label{fig-method-overview}
\Description{The BScout workflow consists of preprocessing and two matching stages. During preprocessing, tree-sitter extracts source functions from source code, and Ghidra decompiles the target binary into functions. Stage 1 maps source and decompiled functions to full-function embeddings, computes their similarity, and retrieves top-N source candidates for each binary function. Stage 2 combines position-driven filtering and supplementation with LLM-assisted selection and review to produce the matching results.}
\end{figure}

We propose \tool{}, a precise function-level binary-to-source SCA method, targeting at matching the binary functions with the corresponding C/C++ source functions. The workflow is presented in Figure~\ref{fig-method-overview}, consisting of three major phases.\space 

Preprocessing leverages Ghidra~\cite{ghidra} to decompile binary files into C/C++-like pseudo functions and uses tree-sitter~\cite{tree-sitter} to parse the source code into functions (the corpus).\space

Stage 1 (\S\ref{sec:stage1}) proposes an embedding-based matching, converting the functions into embedding vectors, measuring the similarity between binary and source functions and retrieving top-$N$ most similar candidates for each binary function. The embedding model is trained based on a large language model with a composition-aware training.\space

Stage 2 (\S\ref{sec:candidate refinement}) utilizes chat-style LLMs and the positional information (i.e., the locations and the function neighborhood) to refine the candidates. It also attempts to uncover the unresolved binary functions as much as possible and finally reports the matching results. 
\space

\subsection{Full-Function Embedding-Based Retrieval and Composition-Aware Training}
\label{sec:stage1}

We take Qwen3.5-4B as the base model to train the embedding model. Using its tokenizer, we observed that real-world C/C++ functions, especially the binary ones, tend to be long. %
The phenomenon leads to a dilemma: truncating the functions and reserving the head tokens may omit a considerable portion of the function code; however, longer token sequences exponentially increase the computing resource and cost for training a large model. To address the issues, we adopt a chunking-based full-function embedding to embed the functions for validation and testing, and a sampling-based variant during the model training.\space

\begin{figure}[!t]
\centering
\includegraphics[width=1\linewidth,height=\textheight,keepaspectratio,clip,trim=0 18 0 0,alt={Panel (a) shows full-function embedding-based retrieval. Source and binary functions are divided into chunks, which are encoded by an embedding model and aggregated into function embeddings. Similarity between the binary-function embedding and source-function embeddings determines the top-N candidates. Panel (b) shows composition-aware training. All-composition loss considers all contributing source functions, while major-composition loss focuses on the major ones. The major-composition loss is weighted by lambda and added to the all-composition loss to train the embedding model.}]{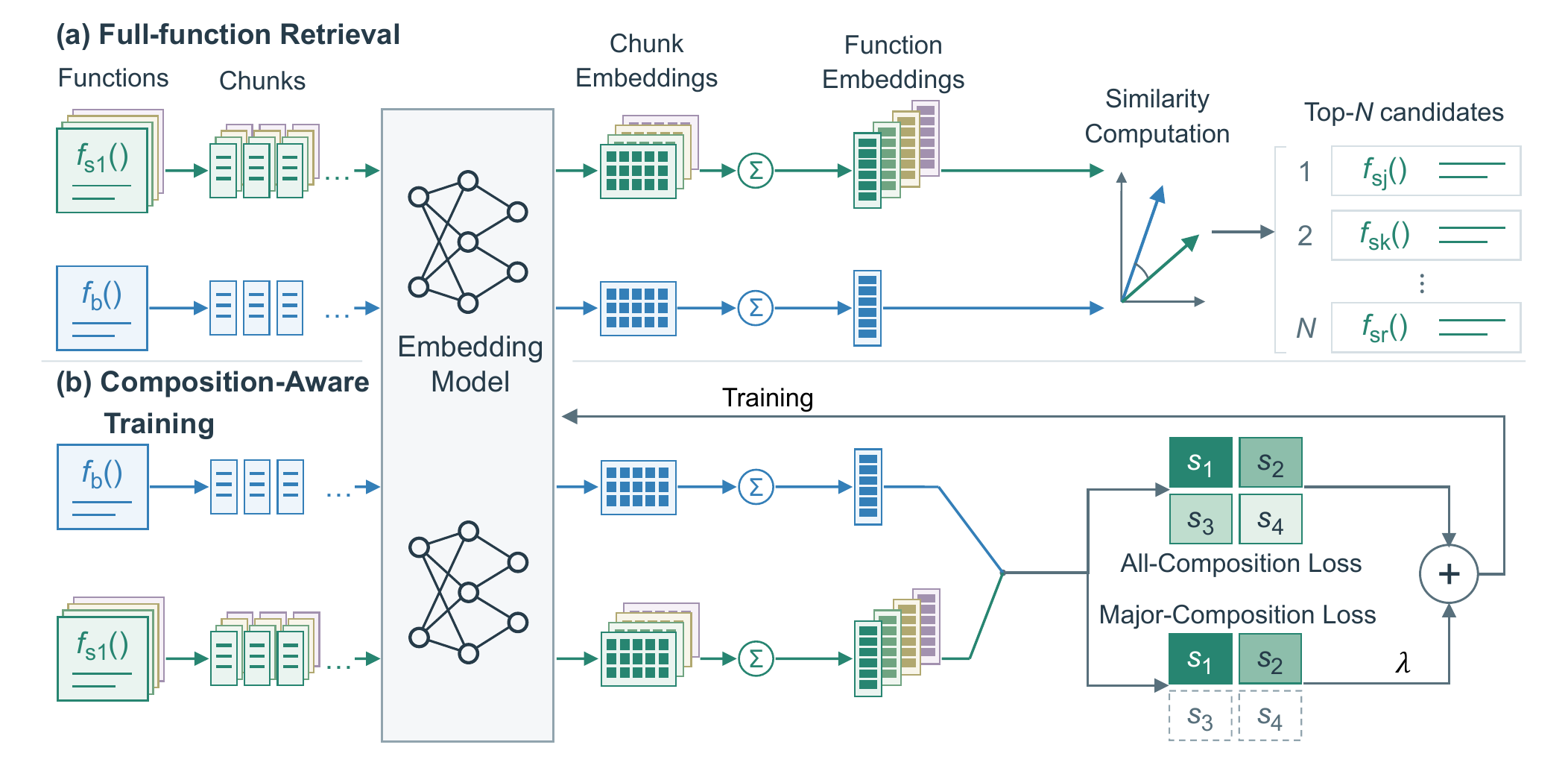}
\caption{
Full-function embedding-based top-$N$ candidates retrieval and composition-aware training.
}\label{fig-hierarchical-supervision}
\Description{Panel (a) shows full-function embedding-based retrieval. Source and binary functions are divided into chunks, which are encoded by an embedding model and aggregated into function embeddings. Similarity between the binary-function embedding and source-function embeddings determines the top-N candidates. Panel (b) shows composition-aware training. All-composition loss considers all contributing source functions, while major-composition loss focuses on the major ones. The major-composition loss is weighted by lambda and added to the all-composition loss to train the embedding model.}
\end{figure}

\subsubsection{Full-Function Embedding and Retrieval} \label{sec:ff-embedding}
Based on the computing resources at hand, for any function $x$, binary or source, we tokenize it into a sequence of tokens and partition the sequence into non-overlapping contiguous chunks. Each chunk is of length 512, with the last one padded if needed. The number of valid tokens, excluding the paddings, is $l_k$ for the chunk $x_k$. The chunks are independently embedded, generating the embedding vector $\mathbf{e}_k$ for $x_k$. The full-function embedding $\mathbf{v}_x$ is then computed as Eq.~\ref{eq:full-function-embedding}. 
\space

\begin{equation}
    \mathbf{v}_x = \frac{\sum_{k=1}^K{l_k \mathbf{e}_k}}{\sum_{k=1}^K{l_k}} 
    \label{eq:full-function-embedding}
\end{equation}

The above method can lead to unacceptable training cost for extremely long functions. To bound the cost, we restrict the number of chunks that a training epoch can observe for a long function. When a function has no more than $m$ chunks, all the chunks are taken into account. But if the number exceeds $m$, we retain the first and last chunks, and randomly sample $m-2$ chunks from the rest. In this paper, we set $m = 8$.\space 

Given the embeddings $\mathbf{b}_i$ and $\mathbf{s}_j$ for a binary function $b_i$ and a source function $s_j$, respectively, their \textit{cosine} similarity, i.e., $\cos(\mathbf{b}_i, \mathbf{s}_j)$, is calculated. Ranking the similarity in a descending order, the top-$N$ most similar candidates (above a predefined threshold, 0.7 in this paper for precision consideration) of $b_i$ can be retrieved for further analysis. \space

\subsubsection{Function Composition-Aware Training}
\label{sec:training}
The composition-aware training attempts to make the embedding model recognize inlined compositions while distinguishing different sources if they make the final binary function with different proportions.
\tool{} scores the multiple source functions of %
$b_i$ by considering both the proportion occupied by each source function $s_{ij}$ and the number of binary instructions $s_{ij}$ hits. Let $L_i$ be the length of $b_i$, $\ell_{ij}$ be the length in $b_i$ tightly surrounding all instructions corresponding to $s_{ij}$, $h_{ij}$ be the hit number, and $\mathcal{S}_i$ be the set of corresponding source functions. Eq.~\ref{eq:composition-score} computes the score $r_{ij}$ and the normalized contribution $\omega_{ij}$ of $s_{ij}$ to $b_i$. The scores for the three source functions in Figure~\ref{fig-background-darktable-multisource} are 0.118, 0.360 and 0.522, respectively.\space

\begin{equation}\label{eq:composition-score}
r_{ij}=\left(\frac{\ell_{ij}}{L_i}\right) h_{ij},
\qquad
\omega_{ij}=\frac{r_{ij}}{\sum_{s_k\in\mathcal{S}_i}r_{ik}}
\end{equation}

In addition to considering all the compositions, we give a further reward to the source functions that form the majority of the binary function. The intuition lies in that, in practice, the majority of a function usually exhibits the semantic behavior of the function. We term such source functions the \textit{major} ones. In this study, we rank the scores in a descending order and $s_{ij}$ is \textit{not} considered major in $b_i$ if any of the following conditions is satisfied: $\omega_{ij} < \frac{1}{|\mathcal{S}_i|}$, $j > 5$, $\frac{\omega_{ij}}{\omega_{i(j-1)}}<0.5$ or $\sum_{1\le k \le n, n < j}\omega_{ik} \ge 0.8$. The major functions of $b_i$ form a set $\mathcal{M}_i$. The scores for the major are re-computed as in Eq~\ref{eq:major-score}.\space

\begin{equation}\label{eq:major-score}
    p_{ij}^{major} = \frac{\omega_{ij}}{\sum_{s_k \in \mathcal{M}_i}\omega_{ik}}
\end{equation}

Let $p_{ij}^{com} = \omega_{ij}$, indicating the score for any composition $s_j$. We adopt contrastive learning to fine-tune Qwen3.5-4B. Eq.~\ref{eq:loss} shows the loss function $\mathcal{L}$ for a batch of $V$ binary-source pairs. The temperature $\tau$ is taken as 0.07~\cite{zhang2025binquery}.\space 

\begin{equation}\label{eq:loss}
    \mathcal{L}^t = -\frac{1}{V}\sum_{i=1}^V \sum_{s_{ij} \in \mathcal{C}_i}p^t_{ij}\log{\frac{\exp(\cos(\mathbf{b}_i, \mathbf{s}_{ij})/\tau)}{\sum_{v=1}^V\sum_{s_{vu} \in \mathcal{C}_v} \exp(\cos(\mathbf{b}_i, \mathbf{s}_{vu})/\tau)}}, 
    \qquad
    \mathcal{L} = \mathcal{L}^{com} + \lambda \mathcal{L}^{major}
\end{equation}

When training the embedding model, a binary-source pair is in the form of $\langle b_i, \mathcal{C}_i \rangle$, where $\mathcal{C}_i$ is $\mathcal{S}_i$ or $\mathcal{M}_i$ depending on $t$ ($com$ or $major$). %
We treat all $s_{ij} \in \mathcal{C}_i$ as positive samples and all $s_{{k\neq i},j} \in \mathcal{C}_{k\neq i}$ as negative samples of $b_i$. The softmax normalization in $\mathcal{L}^t$ expands the composition set $\mathcal{C}$ and computes the similarity between each binary function and each individual source function in the batch, after which the composition-aware weight is applied within the pair. The final loss $\mathcal{L}$ balances the contribution from all the compositions $\mathcal{L}^{com}$ and the major ones $\mathcal{L}^{major}$ with $\lambda$.\space

\subsection{LLM-assisted and Position-driven Candidate Refinement} \label{sec:candidate refinement}

\begin{figure}
\centering
\includegraphics[width=1\linewidth,height=\textheight,keepaspectratio,clip, trim=50 38 50 35,alt={Stage 2 refines the top-N candidates through nine steps: (1) context filtering, (2) candidate selection, (3) anomaly filtering, (4) interval construction, (5) candidate supplementation, (6) LLM as a judge, (7) high-confidence TPL revisiting, (8) candidate selection, and (9) structural filtering. Step 6 includes candidate selection, same-name disambiguation, individual anchor review, n-to-1 match review, and TPL confidence scoring. LLMs support candidate selection at steps 2 and 8 and the decisions at step 6. After step 9, the workflow outputs the matching results or returns to interval construction at step 4 for another round.}]{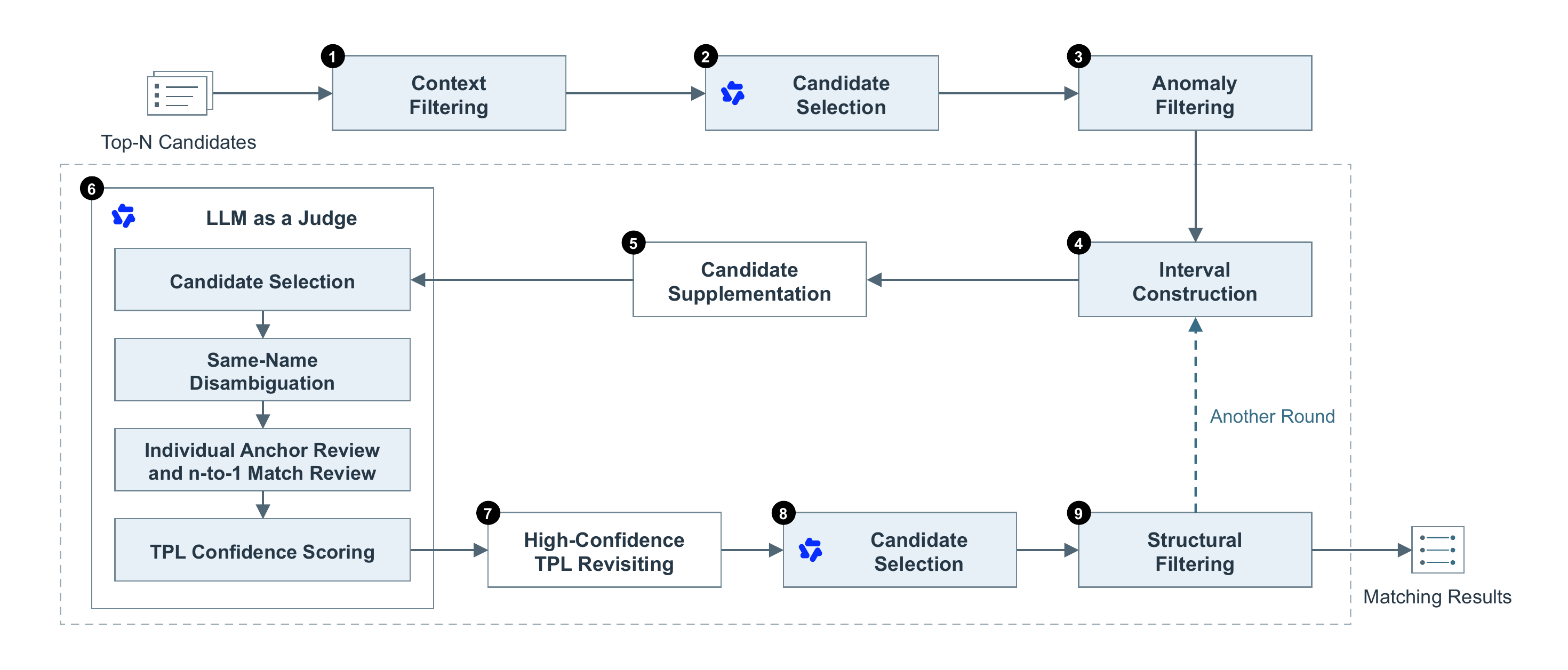}
\caption{Candidate refinement. Blue boxes indicate potential update to the current matches. More rounds can be applied along the dashed arrow, or the current round finalizes the matching results at step \ding{190}.}\label{fig-sccr-overview}
\Description{Stage 2 refines the top-N candidates through nine steps: (1) context filtering, (2) candidate selection, (3) anomaly filtering, (4) interval construction, (5) candidate supplementation, (6) LLM as a judge, (7) high-confidence TPL revisiting, (8) candidate selection, and (9) structural filtering. Step 6 includes candidate selection, same-name disambiguation, individual anchor review, n-to-1 match review, and TPL confidence scoring. LLMs support candidate selection at steps 2 and 8 and the decisions at step 6. After step 9, the workflow outputs the matching results or returns to interval construction at step 4 for another round.}
\end{figure}

Recent LLMs provide the capability of pseudo-source understanding, and the positional information (e.g., the involved source file/TPL and neighbor functions) can indicate whether a match is highly impossible or help find more likely candidates for the unresolved binary functions. 
Therefore, in the candidate refinement stage, an LLM is leveraged to select correct matches from the candidates and to judge the use of specific TPLs, and the positional information helps filter the matches and supplement more candidates. Figure~\ref{fig-sccr-overview} shows the whole process. Starting from the top-$N$ candidates, a list of steps run to emit the final matching results, and multiple rounds can be executed on-demand. Below will discuss the refinement process in detail, despite the orders.\space

\subsubsection{Interval construction}\label{sec:interval-construction}
Interval is the most intuitive representation of positional information, the concept of which is borrowed from~\cite{jiang2024binaryai}. But we give it a redefinition: an \textit{interval} is a range of continuous binary functions starting from and ending with same-file anchors, in the meantime with the number of involved functions constrained by the function number in the corresponding source file. An \textit{anchor} is a binary function that has a determined match. Based on the definition, we can construct intervals (\ding{185}) for given 1-to-1 matching results, after ranking all the anchors and the unresolved (\textit{holes}) by their starting binary address in an ascending order. Overlapped same-file intervals are merged to a bigger one if the result does not exceed the size constraint. The anchors that cannot form or belong to an interval are kept as individual points.\space

Note that, interval construction involves a simple filtering strategy. When an interval contains an individual anchor that is matched to a different TPL with which the interval anchors reside, the match for the anchor is disconnected and the anchor turns into a hole.\space

\subsubsection{Candidate supplementation}\label{sec:supplement}
Supplementing candidates for the unresolved binary functions (\ding{186}) is one of the most important contributions in this paper, which guarantees to discover as many matches as possible. The major concern of this step is to enhance the recall after LLMs potentially reject correct matches from the candidates.\space

We address two kinds of supplementation. First, a hole appears within an interval and it highly probably belongs to the same source file as the interval anchors. Second, a hole outside but near an interval also probably originates from the same file. 
\space
For the first case, we obtain the nearest left and right anchors to the hole of interest and pinpoint their source functions. Then the nearby source functions are scored and the top ones are taken as the candidates of the hole. %
The second case is handled in the same way, except that there is only one nearest anchor, i.e., the left- or right-most in the interval. We expand a fixed window to the left and the right, and within the windows, all the holes are taken into consideration at this step.\space

It must be noted that, no mature approach exists for scoring the source functions. We experimentally propose a solution that concerns the names (a decompiled function may have its name similar to or same as the source function), the callees, the function signatures and the code lengths of the binary and potential source candidate, as well as the position of the candidate around the anchor source functions. In addition, we also take into account whether the potential candidate has ever been selected as a matched candidate $c$ or it is in the same file as $c$. All the above aspects are independently scored and their weighted sum is the score.\space

\subsubsection{LLM as a judge}\label{sec:LLM-as-a-judge}
The LLM attends the following decision-making steps.\space

First, the LLM picks the most confident candidate as the match or rejects all the candidates (\textit{candidate selection} at \ding{183}, \ding{187} and \ding{189}), given a binary function and a set of candidates. %
\space

Second, the LLM performs a further review to three types of resolved matches at \ding{187}. \textit{Same-name disambiguation} attempts to confirm the match of the binary function among a set of same-name source functions {within the same source file as the original match}. \textit{Individual anchor review} double checks the binary-source matches, in which the anchors cannot form a same-file interval with other anchors. \textit{N-to-1 match review} examines the scenario that multiple binary functions match to the same source function, which occurs commonly for the C++ template functions.\space

Third, the LLM decides whether a TPL is really reused by the current binary file (\textit{TPL confidence scoring} at \ding{187}). Based on the matching results so far, \tool{} summarizes the statistics such as the number of matched functions, the number of involved source files, etc. In addition, we empirically weight the factors such as the length of belonging intervals, the position relative to neighboring intervals (for supplemented ones) and so on for the matching results associated to a specific TPL, and select representative samples of the most likely and least likely binary-source pairs. Both the statistics and the selected pairs are fed to the LLM, which emits an answer about whether the TPL is reused: \textit{yes}, \textit{no} or \textit{suspect} that denotes a required further analysis. Along with the answer, the LLM also outputs its confidence, \textit{high}, \textit{medium} or \textit{low}. A filter is applied based on the TPL confidence score. An answer \textit{no} with \textit{high} confidence probably indicates an impossible reuse of the TPL. \tool{} removes all matches associated with the TPL.\space

\subsubsection{Filtering}\label{sec:filtering}
Stage 2 of \tool{} involves three filtering steps besides the aforementioned ones tightly associated with the other steps. 

\textit{Context filtering} (\ding{182}) is applied to the top-$N$ candidates from stage 1. Let $\mathcal{C}_i = \{s_1, s_2, \ldots, s_N\}$ be the candidate set of $b_i$. \tool{} removes $s_j$ from $\mathcal{C}_i$ if $s_j$ locates in a TPL $tpl$, but none of the candidate source functions of $b_k$ around $b_i$ exists in $tpl$. In this study, we inspect all the candidates from $b_{i-5}$ to $b_{i+5}$ (excluding $b_i$) that are continuous in the binary file. 
\space

\textit{Anomaly filtering} (\ding{184}) filters out the matches that connect to the same source function and the number of connections satisfies empirical criteria. Note that, it is also an n-to-1 match case, but $n$ is anomalously large. In \S\ref{sec:implementation} we present the empirical criteria summarized from a real-world project.\space

\textit{Structural filtering} (\ding{190}) finalizes a round and filters the ones newly selected by LLM. It adopts two kinds of filtering. The first is the same as the one mentioned in \S\ref{sec:interval-construction}. The second keeps only newly added matches from \ding{189} if either of the following conditions is satisfied: the binary function in the match can form an interval with existing anchors; the corresponding binary function can form an interval with the others in another newly added match, if (1) the specific TPL has sufficient anchors ($> 2$), or (2) the potential interval involves three or more binary functions and at least two distinct source functions and the embedding-based similarity is always beyond a fixed value ($\ge$ 0.6 in this study) despite the execution round.\space

\subsubsection{High-confidence TPL revisiting}\label{sec:tpl-revisiting} 

The supplementation may still leave functions unresolved, particularly in the areas where anchors are sparse. We revisit the high-confidence TPLs (\ding{188}), seeking more potentially reliable candidates. The behind intuition lies in that, if a TPL is highly confidently reused, it is very likely to have more matches than the results so far, though these potentials may have been suppressed by the similarity threshold or matches from other TPLs. \space

Two kinds of TPLs are considered with high-confidence. First, the LLM reports a \textit{yes} answer with a \textit{high} score for the TPL. Second, the TPL already exhibits a dominant matching result among all TPLs in the dataset, if it has the largest number of matches ($\ge 50$) and the match number accounts for 70\% or more among all the matches and is at least 10 times that of the TPL with the second-largest number of matches. \space

Revisiting a high-confidence TPL recalls the embedding-based matching result only for the TPL of interest. The top-$N$ candidates (over an adaptive threshold $\lambda_h$) of the holes are retrieved for further verification.\space

\section{Evaluation}\label{sec-evaluation}

We implemented a prototype of \tool{} and evaluated it to answer the following three questions. Unless otherwise specified, we present the results of a single round run of \tool{}.\space
\begin{itemize}
    \item \textbf{RQ1:} How effective is \tool{} in binary-to-source function matching? (\S \ref{sec:effectiveness})\space
    \item \textbf{RQ2:} How effective are the proposed techniques in \tool{}? (\S \ref{sec:ablation-stage1} and \S \ref{sec:ablation-stage2})\space
    \item \textbf{RQ3:} What is the efficiency and cost of \tool{}? (\S \ref{sec:efficiency and cost})\space
\end{itemize}

\subsection{Experimental Setup}\label{sec-evaluation-setup}

\subsubsection{Datasets}\label{sec:datasets}
We collected and built multiple datasets for the training, validation and testing. Table~\ref{table:datasets} shows the details and the length statistics.\space

\textbf{Training and Validation.} We downloaded 285 open-source software projects and compiled them using the O2 optimization and keeping the debug information for ground truth construction. All binaries are stripped and later decompiled. In total, we collected 308,160 binary-source pairs. We randomly select 6,000 (6K) as the validation set and the rest are used for model training. As the training on this dataset is time-consuming, we also randomly pick 20,000 pairs (20K) from the big training dataset and train the model on this small dataset to determine the parameter and study the effects of different techniques.\space

\textbf{Testing.} We build two datasets for effectiveness testing. \textit{\binkit{}} is from SBridge~\cite{yang2026sbridge}, which contains 122 programs from the GNU coreutils and inetutils, with 32 compilation configurations (4 architectures $\times$ 2 compilers $\times$ 2 optimizations $\times$ 2 symbol conditions), yielding 3,904 binaries. This dataset contains 39,401 binary-source pairs. \textit{Large-10} contains 10 large open-source projects (including the dependent TPLs) we collected from GitHub. We built it using O2 for x86-64 only, with two variants of binaries, unstripped and stripped. We collected 120 extra open-source projects from GitHub to enlarge the source corpus. The unstripped dataset contains 57,416 pairs and the stripped contains 67,127 pairs.\space

\textbf{Function Characteristics.} From Table~\ref{table:datasets}, we can see that it is common for binary functions to be composed with multiple source functions: more than 1/4 for the training/validation dataset contains and 1/3 for Large-10. BinKit-BCSA is relatively simpler, but near 1/10 binary functions inline other source functions. Concerning the function length, the training and validation dataset exhibits longer binary functions while BinKit-BCSA has generally longer source functions. The testing datasets exhibit a different distribution compared with the training dataset, regarding the lengths of binary functions. 
\space

\begin{table}[t]
\centering
\caption{Datasets. \#GT: number of binary-source pairs (1-to-n); \#SF: involved source functions in GT; MS: ratio of binary functions with inlining. %
The last three columns count the lengths (in tokens) of binary/source functions, in GT (only for training \& validation) or all the functions (for the other datasets).}
\label{table:datasets}
\setlength{\tabcolsep}{3pt}
\begin{tabular}{lll rrr ccc} \toprule
  \multicolumn{2}{l}{\textbf{Datasets}}   &  \textbf{\#GT} & \textbf{\#SF} & \textbf{MS} & \textbf{$>$512} & \textbf{$>$1024} & \textbf{$>$2048} \\
  \midrule
 \multicolumn{2}{l}{Training \& Validation}    & 308,160 & 444,733 & 26.52\% & 49.22\% / \ 8.36\% & 31.12\% / 3.24\% & 16.92\% / 1.30\% \\
 \multicolumn{2}{l}{BinKit-BCSA} & 39,401 & 51,178 & 9.74\% & 17.52\% / 17.24\% & \ 9.38\% / 7.19\% & \ 4.75\% / 1.48\% \\
 \multirow{2}{*}{Large-10} & unstripped & 57,416 & 602,896 & 34.32\% & 28.64\% / \ 7.61\% & 16.86\% / 2.92\% & \ 9.70\% / 0.99\% \\
 & stripped & 67,127 & 654,211 & 34.42\% & 26.52\% / \ 7.61\% & 15.58\% / 2.92\% & \ 8.55\% / 0.99\% \\
 \bottomrule
\end{tabular}
\end{table}

\subsubsection{Implementation} \label{sec:implementation}
\tool{} is implemented in Python, using Qwen3.5-4B and Qwen2Tokenizer as the base model and tokenizer. The training phase runs on a server with three NVIDIA RTX 5880 Ada-48GB GPUs, leverages the gradient cache mechanism~\cite{gao2021gradientcache} and fine tunes in LoRA~\cite{hu2022lora} mode to ensure a batch size as large as possible. The batch size is 384 and 10 epochs are trained, costing about 12 days. The checkpoint with the best performance on the validation dataset is taken for use.\space 

Qwen3.7-Flash (one of the cheapest models) and Qwen3.5-9B (the latest open-source 9B model in the Qwen family) are the backend LLMs in stage 2. The former is accessed through API calling (with thinking mode disabled) and the latter is deployed in the server. To reduce the time cost, we send %
24 and 10 concurrent requests to the Flash and 9B models, respectively. \space

Decompilation is done with Ghidra 11.4 PUBLIC~\cite{ghidra}. And all the analysis steps are conducted on the same server, with 2 $\times$ Intel Xeon Gold 6530 CPUs (128 threads in total) and 512 GB memory.\space

The hyper-parameter $\lambda$ in Eq.~\ref{eq:loss} is 4. We choose a group of values and train models on the 20K dataset. By examining their effect on the validation dataset, we pick 4 that emits the best performance. Stage 1 emits at most top $N$ candidates for each binary function (constrained by a similarity threshold 0.7) and we set $N = 10$ in the experiments. 
\space 

The criteria for anomaly filtering in \S\ref{sec:filtering} draw from our empirical analysis of FreeCAD in Large-10. We set a match number threshold $q$, taking the smaller one between 50 and 0.5\% of the current matched binary-source pairs. Then first, the matches to a source function $s_j$ are eliminated if $s_j$ has $\ge g$ matches where $g = \max(q, 4\times P95, 3\times P98)$. P95 means $s_j$'s match number is in the 95th percentile. Second, after the source functions are ranked by their match numbers in a descending order, we search the first anomalous number $a_k$ satisfying $a_k \ge q$ and $a_k \ge 5\times a_{k+1}$. All matches for the source functions with no less than $a_k$ matches are disconnected. 
The threshold $\lambda_h$ in \S\ref{sec:tpl-revisiting} is lower bounded by 0.5. When a TPL has at least 20 matches, $\lambda_h = \max(0.5, P10 - 0.05)$, with P10 in the 10th percentile of the embedding-based similarity associated with the TPL.\space

\subsubsection{Baselines}
We compare \tool{} against two SOTA methods, BinaryAI~\cite{jiang2024binaryai} and SBridge~\cite{yang2026sbridge}.\space 

BinaryAI is not open-source. But providing a binary file, we can obtain the results from its website, which show the matching results for the binary functions.\space

SBridge natively takes source code as the query and searches the corresponding binary functions. To comply with \tool{} and BinaryAI, we run SBridge without any modification, then take each binary function as the query and rank all the source functions based on their similarity with the query.\space
Because SBridge is super slow on the Large-10 datasets, we did not run it on the extra 120 TPLs. That means, SBridge only queries the source functions within the 10 software projects (including their dependent TPLs). Even reducing the source corpus, SBridge took more than two months to finish the experiment in a parallel manner in the same server, with each binary and the corresponding source code processed on an individual CPU. BinKit-BCSA is much simpler and thus SBridge takes all the 122 programs as the source corpus.\space

\subsubsection{Metrics}
To evaluate the eventual effectiveness of \tool{} and the baselines, we measure their precision ($P$), recall ($R$) and the F1 scores ($=\frac{2\times P \times R}{P + R}$). Since SBridge does not report a confident match or no-match for a binary function as \tool{} and BinaryAI, we use its top-1 candidate as the final result to measure the effectiveness. When inspecting the retrieval effectiveness of stage 1, i.e., the embedding-based matching, we follow what BinaryAI and SBridge measured. We adopt R@$k$ that counts the positive samples detected within the top-$k$ candidates, and MRR ($=\frac{1}{|Q|}\sum_{i=1}^{|Q|} \frac{1}{rank_i}$) that averages the reciprocal ranks among all queries in $Q$.\space 

\begin{table}[tbp]
\centering
\caption{Results on \binkit{}. All numbers are in the form of percentage (\%). \textbf{Bold} ones indicate the best.}\label{tbl-evaluation-public-final-compact}
\centering
\setlength{\tabcolsep}{4pt}
\begin{tabular}{l ccc ccc ccc ccc} \toprule 
  &  \multicolumn{3}{c}{\textbf{BinaryAI} } & \multicolumn{3}{c}{\textbf{SBridge} } & \multicolumn{3}{c}{\textbf{\tool{}-9B} } & \multicolumn{3}{c}{\textbf{\tool{}-Flash} } \\
\cline{2-13}
   & P & R & F1 & P & R & F1 & P & R & F1 & P & R & F1 \\
\hline
\multicolumn{13}{c}{\textit{By architectures}} \\
\hline
ARM32 & 9.51 & 58.24 & 16.34 & 0.25 & 2.68 & 0.45 & \textbf{75.40} & 60.97 & 67.42 & 72.95 & \textbf{83.38} & \textbf{77.82} \\
ARM64 & 9.05 & 56.16 & 15.59 & 0.26 & 2.76 & 0.47 & \textbf{72.06} & 61.45 & 66.33 & 68.76 & \textbf{84.31} & \textbf{75.75} \\
x86 & 9.68 & 58.35 & 16.60 & 0.27 & 2.77 & 0.50 & \textbf{75.88} & 60.20 & 67.13 & 73.32 & \textbf{83.85} & \textbf{78.23} \\
x64 & 9.46 & 59.49 & 16.32 & 0.30 & 2.97 & 0.54 & \textbf{75.94} & 63.01 & 68.87 & 72.27 & \textbf{87.08} & \textbf{78.99} \\
\hline
\multicolumn{13}{c}{\textit{By compilers}} \\
\hline
GCC & 9.32 & 59.41 & 16.12 & 0.29 & 2.99 & 0.53 & \textbf{73.65} & 61.43 & 66.99 & 70.00 & \textbf{85.16} & \textbf{76.84} \\
Clang & 9.52 & 56.69 & 16.30 & 0.25 & 2.60 & 0.45 & \textbf{75.95} & 61.40 & 67.91 & 73.65 & \textbf{84.17} & \textbf{78.56} \\
\hline
\multicolumn{13}{c}{\textit{By optimization levels}} \\
\hline
O0 & 9.96 & 65.03 & 17.28 & 0.31 & 2.92 & 0.56 & \textbf{76.16} & 69.90 & 72.90 & 74.14 & \textbf{90.00} & \textbf{81.31} \\
O2 & 8.48 & 47.72 & 14.40 & 0.22 & 2.61 & 0.41 & \textbf{71.99} & 48.84 & 58.20 & 67.97 & \textbf{76.75} & \textbf{72.10} \\
\hline
\multicolumn{13}{c}{\textit{By symbol status}} \\
\hline
unstripped & 9.39 & 60.97 & 16.27 & 0.34 & 3.58 & 0.62 & \textbf{73.54} & 79.23 & 76.28 & 71.33 & \textbf{93.51} & \textbf{80.93} \\
stripped & 9.45 & 55.07 & 16.13 & 0.20 & 1.99 & 0.36 & \textbf{77.21} & 43.17 & 55.38 & 72.32 & \textbf{75.60} & \textbf{73.93} \\
\hline \hline
\textbf{Overall} & 9.42 & 58.06 & 16.21 & 0.27 & 2.79 & 0.49 & \textbf{74.77} & 61.42 & 67.44 & 71.76 & \textbf{84.67} & \textbf{77.68} \\
\bottomrule
\end{tabular}
\end{table}

\begin{table}[!htbp]
\centering
\caption{Results on Large-10. All numbers are in the form of percentage (\%).}\label{tbl-evaluation-large10-final}

\setlength{\tabcolsep}{3pt}
\begin{tabular}{cl ccc ccc ccc ccc}
\toprule
 & & \multicolumn{3}{c}{\textbf{BinaryAI}} & \multicolumn{3}{c}{\textbf{SBridge}} & \multicolumn{3}{c}{\textbf{\tool{}-9B}} & \multicolumn{3}{c}{\textbf{\tool{}-Flash}} \\
\cline{3-14}
    &    & P & R & F1 & P & R & F1 & P & R & F1 & P & R & F1 \\
\midrule
\multirow{11}{*}{\rotatebox{90}{\textit{Unstripped}}} &
EncryptPad & 62.46 & 31.62 & 41.99 & 0.00 & 0.00 & 0.00 & \textbf{75.03} & 61.40 & 67.53 & 70.68 & \textbf{72.16} & \textbf{71.41} \\
&darktable & 32.93 & 16.31 & 21.81 & 0.20 & 0.12 & 0.15 & \textbf{98.68} & 73.94 & 84.54 & 96.88 & \textbf{82.05} & \textbf{88.85} \\
&HeadPhone & 81.09 & 43.52 & 56.64 & 0.00 & 0.00 & 0.00 & \textbf{82.26} & 66.40 & 73.49 & 77.35 & \textbf{72.38} & \textbf{74.78} \\
&Audacity & 66.69 & 12.26 & 20.72 & 0.00 & 0.00 & 0.00 & \textbf{98.85} & 25.05 & 39.97 & 95.71 & \textbf{32.38} & \textbf{48.39} \\
&FreeCAD & 12.82 & 1.69 & 2.99 & 0.00 & 0.00 & 0.00 & \textbf{94.09} & 28.23 & 43.43 & 83.91 & \textbf{35.95} & \textbf{50.34} \\
&NotepadNext & 47.10 & 14.86 & 22.59 & 0.00 & 0.00 & 0.00 & 84.23 & 51.81 & 64.15 & \textbf{85.97} & \textbf{65.54} & \textbf{74.38} \\
&LMMS & 23.44 & 3.77 & 6.49 & 0.00 & 0.00 & 0.00 & \textbf{99.54} & 31.62 & 47.99 & 98.15 & \textbf{43.22} & \textbf{60.01} \\
&Notes & 31.99 & 5.61 & 9.55 & 0.00 & 0.00 & 0.00 & \textbf{99.49} & 34.16 & 50.86 & 99.02 & \textbf{40.34} & \textbf{57.33} \\
&Transmission & 23.27 & 6.35 & 9.98 & 0.00 & 0.00 & 0.00 & \textbf{97.94} & 42.73 & 59.50 & 95.84 & \textbf{54.57} & \textbf{69.54} \\
&Clementine & 90.30 & 17.60 & 29.45 & 0.03 & 0.01 & 0.02 & \textbf{99.24} & 36.82 & 53.71 & 97.70 & \textbf{51.50} & \textbf{67.45} \\
\cline{2-14}
&\textbf{Overall} & 52.60 & 12.35 & 20.01 & 0.02 & 0.01 & 0.02 & \textbf{95.08} & 38.37 & 54.68 & 93.08 & \textbf{48.93} & \textbf{64.14} \\
\midrule
\multirow{11}{*}{\rotatebox{90}{\textit{Stripped}}} &
EncryptPad & 63.70 & 43.52 & 51.71 & 0.00 & 0.00 & 0.00 & \textbf{76.02} & 16.80 & 27.52 & 61.85 & \textbf{51.18} & \textbf{56.01} \\
&darktable & 34.66 & 15.39 & 21.32 & 0.12 & 0.11 & 0.12 & \textbf{96.72} & 55.10 & 70.20 & 94.18 & \textbf{67.19} & \textbf{78.43} \\
&HeadPhone & \textbf{88.24} & 43.80 & 58.54 & 0.00 & 0.00 & 0.00 & 80.42 & 59.54 & 68.42 & 77.90 & \textbf{65.07} & \textbf{70.91} \\
&Audacity & 69.00 & 9.74 & 17.06 & 0.00 & 0.00 & 0.00 & \textbf{96.54} & 18.78 & 31.44 & 90.00 & \textbf{25.70} & \textbf{39.98} \\
&FreeCAD & 13.29 & 1.58 & 2.83 & 0.00 & 0.00 & 0.00 & \textbf{96.19} & 24.53 & 39.09 & 82.09 & \textbf{31.86} & \textbf{45.90} \\
&NotepadNext & 47.84 & 13.09 & 20.56 & 0.00 & 0.00 & 0.00 & \textbf{82.27} & 4.21 & 8.01 & 71.74 & \textbf{24.37} & \textbf{36.38} \\
&LMMS & 23.49 & 3.37 & 5.89 & 0.00 & 0.00 & 0.00 & \textbf{99.28} & 27.99 & 43.67 & 97.65 & \textbf{38.96} & \textbf{55.70} \\
&Notes & 30.03 & 3.42 & 6.13 & 0.00 & 0.00 & 0.00 & \textbf{94.04} & 6.09 & 11.44 & 84.01 & \textbf{17.17} & \textbf{28.52} \\
&Transmission & 23.38 & 4.35 & 7.34 & 0.00 & 0.00 & 0.00 & \textbf{83.41} & 3.98 & 7.59 & 74.75 & \textbf{16.53} & \textbf{27.07} \\
&Clementine & 90.67 & 14.57 & 25.11 & 0.01 & 0.01 & 0.01 & \textbf{96.58} & 6.96 & 12.98 & 79.62 & \textbf{21.90} & \textbf{34.36} \\
\cline{2-14}
&\textbf{Overall} & 53.54 & 10.34 & 17.33 & 0.01 & 0.01 & 0.01 & \textbf{95.25} & 18.24 & 30.62 & 85.54 & \textbf{29.98} & \textbf{44.40} \\
\bottomrule
\end{tabular}
\end{table}

\subsection{Overall Effectiveness} \label{sec:effectiveness}
Table~\ref{tbl-evaluation-public-final-compact} and Table~\ref{tbl-evaluation-large10-final} present the results for the tools over BinKit-BCSA and Large-10, respectively. \tool{} with different backend LLMs are distinguished with the postfixes. \space

Following the presentation in the SBridge paper~\cite{yang2026sbridge}, we show the details by the architectures, the compilers, the optimization levels and the symbol status, for \binkit{} in Table~\ref{tbl-evaluation-public-final-compact}. At the end, we show an overall performance. On this simpler dataset, \tool{} completely outperforms both BinaryAI and SBridge regarding the precision and F1, using either the 9B model or the bigger Flash model. For the \textit{stripped}, \tool{}-9B shows lower recall than BinaryAI, but its much higher precision (77.21\% \textit{vs.} 9.45\%) results in a high comprehensive performance. Eventually, \tool{}-9B achieves the best precision and \tool{}-Flash has the highest recall and F1, in each category and over the whole dataset. \space

On Large-10, we separate the results for unstripped and stripped binaries and show the result for each target software project. SBridge nearly failed on every software project to pick the corresponding source function as top-1 candidate of any binary function. BinaryAI is much better and even achieves higher precision or recall in a few cases than \tool{}. But \tool{}-Flash achieves the best recall in all cases and \tool{}-9B shows the best precision in most cases. \space
Clearly, in either dataset, \tool{}-9B can provide more precise matching results than \tool{}-Flash because it reports much fewer matches, especially for the stripped binaries. In fact, a smaller model is more likely to fail to identify the correct candidate within the top $N$, thus reducing reliable anchors and intervals. As a result, the supplementation step is affected and even if a correct match enters the supplemented candidates, it is probably rejected by the 9B model. 
\space

In summary, \tool{} has the overall performance, using either backend LLM, surpassing the baselines and is demonstrated to be effective.\space

\subsection{Effectiveness of Full-Function Retrieval and Composition-Aware Training} \label{sec:ablation-stage1}

In this section, we evaluate the effectiveness of the proposed techniques in stage 1. %
Because training the model on the big dataset is time-consuming, we train each variant (8 in total as in Table~\ref{tbl-evaluation-rq2-ablation}) on the 20K dataset (\S\ref{sec:datasets}) for 5 epochs (each training takes 6 $\thicksim$ 25 hours), choose the best checkpoint on the validation set and evaluate it on 6,000 randomly selected pairs (same size as the validation dataset) from Large-10-stripped. \space

\begin{table}[!t]
\centering
\caption{Embedding-based retrieval with different configurations. (R@10) is for multi-source functions.}\label{tbl-evaluation-rq2-ablation}
\begin{tabular}{r l r r r r}
\toprule
\textbf{ID}& {\textbf{Configurations}} & { \textbf{R@1(\%)}} & {\textbf{R@5(\%)}} & { \textbf{R@10(\%)}} & { \textbf{MRR}} \\
\midrule
1 & {\raggedright Trunc-512 / Entry-only\par} & {\raggedleft 11.40\par} & {\raggedleft 21.83\par} & {\raggedleft 28.62(39.77)\par} & {\raggedleft 0.1697\par}  \\
2 & {\raggedright Trunc-512 / Composition+Major\par} & {\raggedleft 12.22\par} & {\raggedleft 24.37\par} & {\raggedleft 30.53(43.58)\par} & {\raggedleft 0.1860\par}  \\
3 & {\raggedright Trunc-2048 / Entry-only\par} & {\raggedleft 12.28\par} & {\raggedleft 23.75\par} & {\raggedleft 30.07(45.42)\par} & {\raggedleft 0.1836\par} \\
4 & {\raggedright Trunc-2048 / Composition+Major\par} & {\raggedleft 14.65\par} & {\raggedleft 27.90\par} & {\raggedleft 34.07(46.55)\par} & {\raggedleft 0.2145\par} \\
5 & {\raggedright Full-Function / Entry-only\par} & {\raggedleft 11.53\par} & {\raggedleft 23.22\par} & {\raggedleft 30.05(44.41)\par} & {\raggedleft 0.1737\par}  \\
6 & {\raggedright \textbf{Full-Function / Composition+Major}\par} & {\raggedleft \textbf{16.02}\par} & {\raggedleft \textbf{29.15}\par} & {\raggedleft \textbf{37.27}(\textbf{55.47})\par} & {\raggedleft \textbf{0.2313}\par}  \\
\hline
7 & {\raggedright Full-Function / Composition-only\par} & {\raggedleft 13.73\par} & {\raggedleft 26.05\par} & {\raggedleft 34.18(53.75)\par} & {\raggedleft 0.2035\par}  \\
8 & {\raggedright Full-Function / Major-only\par} & {\raggedleft 12.85\par} & {\raggedleft 24.70\par} & {\raggedleft 31.27(50.24)\par} & {\raggedleft 0.1895\par}  \\

\bottomrule
\end{tabular}
\end{table}

\begin{table}[t]
    \centering
    \caption{Results for at least one/two compositions are in R@10 (\%) for multi-source binary functions.}
    \label{tbl:r10-multi-source}
    \begin{tabular}{l r r r} 
    \toprule
      \textbf{Dataset}   & \textbf{BinaryAI} & \textbf{SBridge} & \textbf{\tool{}} \\
      \midrule
      \binkit{}         &  80.38 / 3.86  & 14.62 / 3.23   &  \textbf{96.25} /  \textbf{42.97}  \\
      Large-10-unstripped & 23.53 /  0.23 & 0.08 / 0.00 &   \textbf{64.58} /  \textbf{10.64} \\
      Large-10-stripped & 20.59 / 0.15 & 0.08 / 0.00 &  \textbf{57.32} /  \ \ \textbf{9.02}
      \\
      \bottomrule
    \end{tabular}
\end{table}

There are three embedding strategies, i.e., truncating (\textit{Trunc}) the first 512 tokens for each function, truncating the first 2,048 tokens and adopting full-function embedding as in \S\ref{sec:ff-embedding}. Two pairing strategies are provided, i.e., 1-to-1 pairing between a binary function and its corresponding entry source function (e.g., the first one in Figure~\ref{fig-background-darktable-multisource}), and 1-to-n multi-source pairing, corresponding to traditional contrastive learning loss function and the Eq.~\ref{eq:loss} loss function, respectively, for each embedding strategy.\space
Full-function embedding with all compositions and the major ones achieves the best performance among the six variants (\#1 $\thicksim$ \#6). For the multi-source function-matches, the approach improves R@10 by at least 8.92\% (46.55\% $\rightarrow$ 55.47\%).\space

We also separately train a model with full-function embedding but using $\mathcal{L}^{com}$ or $\mathcal{L}^{major}$ in Eq.~\ref{eq:loss}, i.e., \textit{composition-only} (\#7) and \textit{major-only} (\#8) in Table~\ref{tbl-evaluation-rq2-ablation}. It shows that the all-composition factor is more effective than the major-composition factor, but combining both produces a better result.\space

Furthermore, by examining how many multi-source functions have their compositions fall into the top-10, we can see that \tool{} significantly outperforms the competitors in Table~\ref{tbl:r10-multi-source}. In particular, 10 times ($\frac{42.97}{3.86}$) more functions have at least two compositions within top-10 by \tool{} compared with BinaryAI.\space

The results demonstrate that the full-function embedding and the proposed composition-aware training are both effective in improving the performance of the embedding-based binary-source function retrieval, with a particular augmentation for the multi-source scenario. \space

\begin{table}[t]
\centering
\caption{Ablation study for the two-stage approach and two important technical strategies in stage 2.}\label{tbl-evaluation-fcsmatch-stage-comparison}
\begin{tabular}{l l r r r}
\toprule
\textbf{Dataset} & \textbf{Configuration} & \textbf{P (\%)} & \textbf{R (\%)} & \textbf{F1 (\%)} \\
\midrule
\multirow{4}{*}{BinKit-BCSA} & \tool{}-Flash & \textbf{71.76} & \textbf{84.67} & \textbf{77.68} \\
 & \hspace{1em} w/o Stage 2 & 35.30 & 78.84 & 48.76 \\
 & \hspace{1em} w/o LLM & 35.82 & 81.44 & 49.76 \\
 & \hspace{1em} w/o Supplement & 68.09 & 75.88 & 71.77 \\
 \hline
\multirow{4}{*}{Large-10-unstripped}  & \tool{}-Flash & \textbf{93.08} & \textbf{48.93} & \textbf{64.14} \\
 & \hspace{1em} w/o Stage 2 & 41.71 & 38.11 & 39.83 \\
 & \hspace{1em} w/o LLM & 40.67 & 43.13 & 41.86 \\
 & \hspace{1em} w/o Supplement & 87.74 & 40.51 & 55.43 \\
 \hline
\multirow{4}{*}{Large-10-stripped}  & \tool{}-Flash & \textbf{85.54} & \textbf{29.98} & \textbf{44.40} \\
 & \hspace{1em} w/o Stage 2 & 29.71 & 28.45 & 29.07 \\
 & \hspace{1em} w/o LLM & 25.20 & 29.89 & 27.34 \\
 & \hspace{1em} w/o Supplement & 75.69 & 24.57 & 37.10 \\
\bottomrule
\end{tabular}
\end{table}

\begin{table}[t]
\centering
\caption{Contribution of Stage 2 to the final matching results. \textit{\#Final} shows final reports by \tool{}, and \textit{\#Newly Found} indicates the matches discovered solely by stage 2 and the number/ratio of correct matches.}\label{tbl-contribution-stage2}
\begin{tabular}{l r c}
\toprule
\textbf{Dataset} & \textbf{\#Final} & \textbf{\#Newly Found} \\
\midrule
BinKit-BCSA & {46,484} & 4,400  / 2,813(63.93\%)  \\
Large-10-unstripped & {30,181}& 4,781  / 4,426(92.57\%)  \\
Large-10-stripped & {23,528} & 3,485  / 2,793(80.14\%)  \\
\bottomrule
\end{tabular}
\end{table}

\subsection{Effectiveness of Candidate Refinement} \label{sec:ablation-stage2}

In this section, we evaluate the effectiveness of stage 2, including how stage 2 improves the results from stage 1 and how the adopted refinement techniques affect the results. As \tool{}-Flash exposes generally higher F1 scores in \S\ref{sec:effectiveness}, we perform the analysis focusing on it in this section.\space

\subsubsection{Necessity of Stage 2} \label{sec:ablation-necessity-stage2}
We first check the effectiveness of stage 2 (running one round). In Table~\ref{tbl-evaluation-fcsmatch-stage-comparison}, the configuration \textit{w/o Stage 2} takes the top-1 candidates from stage 1 as the matching results. It is obvious that stage 2 in \tool{}-Flash greatly improves the precision and simultaneously enhances the recall, demonstrating the necessity of stage 2.\space

We further examine how many matches are definitely uncovered by stage 2, i.e., counting the ones not within the top 10 candidates in stage 1. From Table~\ref{tbl-contribution-stage2}, we can see that about 1/10 and 1/7 are newly found by stage 2 for \binkit{} and Large-10, respectively. Among them, most ($>$ 60\%) are finally identified as correct matches.\space

Based on the above discussion, stage 2 is significant to improve the performance of \tool{}.\space

\subsubsection{Ablation Study of Two Essential  Techniques in Stage 2}\label{sec:ablation-two-techniques-in-stage2}
In stage 2, the two most important techniques are the adoption of LLM as a judge and supplementing candidates for unresolved functions. We remove either from \tool{} to validate the corresponding effect. The results are shown in Table~\ref{tbl-evaluation-fcsmatch-stage-comparison}.\space

Removing LLM (\textit{w/o LLM}), we use the top-1 similar candidate as a confident match whenever the LLM is required to pick one from a group of candidates for an unresolved binary function. High-confidence TPLs are those with at least 20 anchors. The review of the resolved matches in \S\ref{sec:LLM-as-a-judge} is canceled. The precision drops significantly while the recall changes relatively slightly for all datasets. Disabling candidate supplementation (\textit{w/o Supplement}) will terminate the subsequent steps but keep all preceding steps. In Table~\ref{tbl-evaluation-fcsmatch-stage-comparison}, canceling the supplement (and the subsequent steps) reduces both precision and recall. More specifically, the newly found matches in Table~\ref{tbl-contribution-stage2} will be missed. But compared with removing LLM, the precision is relatively high, causing higher F1 scores. \space 

Therefore, we can conclude that the adoption of LLM mainly contributes to emitting precise binary-source function matches and the supplementation helps uncover more matches.\space

\subsubsection{Effect of Each Update Step in Stage 2}
\label{sec:ablation-steps-stage2}

\begin{table}[t]
\centering
\caption{Effects and costs at each single update step on Large-10-stripped.}\label{tbl-effects-costs}
\begin{tabular}{crl rrrr}
\toprule
& \multicolumn{2}{l}{\textbf{Update Step}} & \textbf{P (\%)} & \textbf{R (\%)} & \textbf{F1 (\%)} & \textbf{API Cost (USD)} \\
\midrule
\multirow{3}{*}{\rotatebox[origin=c]{90}{Initial}} &
\ding{182} & Context Filtering & 30.82 & 28.52 & 29.62 & N/A \\
& \ding{183} & Candidate Selection  & 71.59 & 25.32 & 37.41 & 10.19 \\
& \ding{184} & Anomaly Filtering & 75.69 & 24.57 & 37.10 & N/A \\
\midrule
\multirow{7}{*}{\rotatebox[origin=c]{90}{Round 1}} &
\ding{185} & Interval Construction & 81.53 & 24.23 & 37.35 & N/A \\
& \ding{187} & Candidate Selection & 80.85 & 29.30 & 43.01 & 19.36 \\
& & Same-Name Disambiguation & 82.07 & 29.02 & 42.88 & 1.03 \\
& & Individual Anchor \& n-to-1 Match Review & 87.48 & 27.95 & 42.36 & 1.40 \\
& & TPL Confidence Scoring & 87.95 & 27.95 & 42.42 & 0.09 \\
& \ding{189} & Candidate Selection  & 82.22 & 30.46 & 44.45 & 17.06 \\
& \ding{190} & Structural Filtering & 85.54 & 29.98 & 44.40 & N/A \\
\midrule
\multirow{4}{*}{\rotatebox[origin=c]{90}{Round 2}} &
\ding{187} & Candidate Selection & 81.71 & 32.19 & 46.18 & 59.24 \\
& \ding{187} & LLM as a Judge & 85.14 & 30.95 & 45.40 & 2.69 \\
& \ding{189} & Candidate Selection & 80.82 & 32.03 & 45.88 & 17.92 \\
& \ding{190} & Structural Filtering & 84.16 & 31.55 & 45.90 & N/A \\
\midrule
& \multicolumn{2}{l}{\textbf{Total}} & -- & -- & -- & 128.98 \\
\bottomrule
\end{tabular}
\end{table}

We further examine how each update step (the blue boxes in Figure~\ref{fig-sccr-overview}) within the first round affects the results. The \textit{Initial} and \textit{Round 1} parts in Table~\ref{tbl-effects-costs} report the matching performance after each update for \tool{}-Flash on Large-10-stripped. For context filtering (\ding{182}), we take the top-1 as done in \S\ref{sec:ablation-necessity-stage2}, while for the other steps, an anchor indicates a match.\space

The first adoption of LLM (\ding{183}) dramatically improves the precision (30.82\% $\rightarrow$ 71.59\%) with declined recall but enhanced F1 ($\uparrow$7.78\%). The first interval construction (\ding{185}) and the involved simple filtering contribute the second highest precision gain ($\uparrow$5.84\%) but show the lowest recall in Table~\ref{tbl-effects-costs}. The largest recall gain benefits from the first candidate supplementation, which can be seen at \ding{187} in round 1 ($\uparrow$5.07\%). Each single step after candidate selection in \ding{187} increases the precision until reaching the peak, while generally reducing the recall. But after addressing the high-confidence TPLs, the recall and F1 reach the highest (\ding{189} in round 1).\space

Clearly, each update step contributes to only one of precision and recall after the previous step. \tool{} users can decide to trade off and terminate at a certain step.\space

\subsubsection{Effect of More Rounds}
\label{sec:ablation-more-rounds}

We examine the benefit and cost of running a second round. In Table~\ref{tbl-effects-costs}, round 1 ends with precision 85.54\% and recall 29.98\% and we collapse the steps in \ding{187} except candidate selection. The metric values in round 2 fluctuate as the pipeline advances. The recall is always higher than in round 1 but the precision is lower than at the end of round 1. Fortunately, we get higher F1 scores.\space

Note that the API cost of round 2 is higher, twice that for round 1. Compared to the cost increment of running one more round, the performance gain is quite marginal. \space

\subsection{Efficiency and Cost}\label{sec:efficiency and cost}

We evaluate the efficiency and cost of \tool{}-Flash for a single refinement round %
in Table~\ref{tbl-evaluation-system-cost}.\space

\binkit{} has a considerable number of binaries, taking much more time to finish a complete round of matching, including decompilation and embedding. Stage 1 matching takes only 0.45 seconds per binary file on average as the source corpus is relatively small. Large-10 contains fewer but larger binaries and a huge source corpus, thus taking less time for decompilation and embedding on binaries but more average time for stage 1 matching (approximately 40 seconds to 1 minute). For stage 2, 15 seconds, 24 minutes and 35 minutes are spent per binary for the three datasets, and a non-negligible portion of the time cost is consumed by the non-LLM analysis, from 21.6\% to 47.0\%. The time cost is acceptable. Even for stripped software in Large-10, the average total time is 50 minutes and the unstable part is the stage 1 matching, which is linear with the size of the source corpus, while the others are generally bounded by the size of the binary.\space

The API calls to the model, the consumed tokens and the API cost are also bounded by the binary size. As a result, each binary in \binkit{} costs much less than those in Large-10. 
On average, the API cost per binary is as low as about 0.01 USD on \binkit{} and approximately 3 $\thicksim$ 5 USD on Large-10. \space

The results make \tool{} practical for binary-source function matching of real-world large software on a large-scale source corpus.\space

\begin{table}[tbp]
\centering
\caption{Efficiency and monetary cost for \tool{}-Flash. \#B: number of binary files; Dec, Emb and Stage 1 Matching: time cost for decompilation, binary function embedding and retrieval, in hours; Stage 2: time cost (non-LLM analysis time); \#APIs: total number of API calls to the LLM; Tokens: in billions; Cost: in USD. }\label{tbl-evaluation-system-cost}
\setlength{\tabcolsep}{4pt}
\begin{tabular}{l r rrr rrr r}
\toprule
\textbf{Dataset} & \textbf{\#B} &  \textbf{Dec} & \textbf{Emb} & \makecell[r]{\textbf{Stage 1}\\\textbf{Matching}} & \textbf{Stage 2} & \textbf{\#APIs} & \textbf{Tokens} & \textbf{Cost}  \\
\midrule
\binkit{}           & 3,904& 15.46 & 7.51 & 0.49 & 16.80 (3.63) & 175,933 & 1.237 & \$43.92 \\
Large-10-unstripped & 10   & 1.92 & 0.87 & 0.11 & 3.96 (1.86) & 123,287 & 0.826 & \$31.18  \\
Large-10-stripped   & 10   & 1.23 & 1.05 & 0.16 & 5.82 (1.62) & 178,513 & 1.323 & \$49.13 \\
\bottomrule
\end{tabular}
\end{table}

\section{Discussion}\label{sec-discussion-threats}

We further discuss the design choices, practical trade-offs, and possible improvements of \tool{}.

\textbf{Function representation and composition supervision.} \tool{} combines full-function embedding with composition-aware training to account for both long functions and the multiple source correspondences introduced by inlining. The results in \S\ref{sec:ablation-stage1} support this design. However, the chunks of a function are encoded independently before their embeddings are aggregated. Relationships between operations in different chunks are therefore not explicitly modeled. This may limit the representation of functions. %
It is actually a trade-off due to limited computing resources at hand. With sufficient GPUs, full-function embedding can be done in one go, and a real full-function embedding can be also adopted for the training phase. %

\textbf{Position-driven candidate supplementation.} Candidate supplementation allows \tool{} to search beyond the initial top-$N$ candidates, but it relies on the source-file and TPL clues provided by existing matches. Sparse anchors may leave insufficient positional evidence to identify additional candidates. Revisiting high-confidence TPLs provides another search opportunity in these areas, although the TPLs considered are still determined from existing matching evidence. Some functions may therefore remain unresolved when reliable evidence for their originating TPL is insufficient. Moreover, supplemented candidates are not necessarily correct matches and still require LLM selection and subsequent filtering. The results in \S\ref{sec:ablation-stage2} support combining these steps, while also showing the trade-off between recovering more matches and rejecting unreliable ones.

\textbf{Backend LLM and refinement rounds.} Different models can be selected as the backend according to practical requirements. SOTA LLMs generally provide better overall performance, with advantages in precision, recall, or their combined effectiveness. Smaller, locally deployed models may perform somewhat less well overall, but can still achieve useful results and have practical value. More advanced LLMs may therefore be preferred when matching quality is the primary concern, while locally deployed smaller models remain viable choices under tighter constraints on cost, deployment, or data privacy. Additional refinement rounds also involve a trade-off between performance and overhead. On Large-10-stripped, a second round of \tool{}-Flash improves recall and F1 but reduces precision, with substantial additional API cost (see \S\ref{sec:ablation-more-rounds}). In practice, the number of refinement rounds should therefore be assessed against the task objectives, performance requirements, and acceptable computational or API cost, rather than fixed in advance.

\textbf{LLM-assisted refinement of decompiled code.} The examples in \S\ref{sec-background-motivation} show how compilation and inlining can change the calls and expressions of a function. We explored using ChatGPT, DeepSeek and other LLMs to refine decompiled functions. In some cases, the rewritten code was nearly identical in form to the corresponding source code. This observation suggests a possible use of LLMs in input preparation before embedding-based retrieval. Such refinement may reduce differences in code presentation in certain cases, but we have also observed significant difference between the refined decompiled code and the original source code in other cases. The effect on both the refinement and the retrieval needs to be further evaluated. %

\textbf{Threats to validity.} Our evaluation is subject to the coverage of both the ground truth and the test datasets. If a valid correspondence is absent from the ground truth and a predicted match to it is counted as incorrect, both precision and recall are underestimated. %
The present evaluation does not quantify the possible effects. Dataset coverage also limits generalization. \binkit{} includes multiple architectures, compilers, optimization levels, and symbol conditions, but it contains only the  small utility programs. %
Large-10 includes larger software projects, but is built only for x86-64 with O2. The two datasets provide complementary coverage; performance on other programs and build configurations requires further evaluation.

\section{Related Work}\label{sec-related-work}

\subsection{Binary-to-Source Function Matching and Software Composition Analysis}\label{sec-related-work-sca}

Compilation changes the form and structure of source code, making binary-to-source function matching difficult. BinPro~\cite{miyani2017binpro} and B2SFinder~\cite{feng2019b2sfinder} use traceable features, including constants, library calls, function-call relations, and project structure, to establish correspondences between binaries and source code. Beyond these features, CodeCMR~\cite{yu2020codecmr} learns a shared embedding space for binary and source functions. BinaryAI~\cite{jiang2024binaryai} combines function-vector retrieval with link-time locality and call relations. SBridge~\cite{yang2026sbridge} compares control blocks across the two domains to handle differences introduced by compilation and inlining.

Software composition analysis also studies reuse at the library and component levels. LibDB~\cite{tang2022libdb} combines function features and call graphs, while ModX~\cite{yang2022modx} and LibAM~\cite{li2023libam} match functional modules and call-graph regions, respectively. BinCoFer~\cite{zou2025bincofer} and DeRed~\cite{chang2026dered} address redundant features and deceptive reuse. Wang et al.~\cite{wang2024gap} examine how reference-library build configurations affect binary software composition analysis. Other work studies modified reuse~\cite{woo2021centris}, component dependencies~\cite{jiang2023tplite,na2024cneps,xu2026deepsca}, supply-chain vulnerabilities~\cite{hu2026vulsca}, and library versions~\cite{choi2025tiver,dong2024libvdiff}.

\subsection{Function Representations and Inlining-Aware Matching}\label{sec-related-work-representation}

Function representations capture control flow, dependencies, and instruction semantics in different ways. jTrans~\cite{wang2022jtrans} models jumps, while CodeArt~\cite{su2024codeart} incorporates program dependencies. DiEmph~\cite{xu2023diemph} reduces biases caused by instruction distributions. HermesSim~\cite{he2024hermessim}, vSim~\cite{wang2026vsim}, and VexIR2Vec~\cite{venkatakeerthy2025vexir2vec} use semantic graphs, value semantics, and intermediate representations, respectively. Nova~\cite{jiang2025nova} combines hierarchical attention with contrastive learning. For long code, LongCoder~\cite{guo2023longcoder} extends the attention context, while SEA~\cite{hu2024sea} encodes code chunks separately and aggregates their vectors.

Inlining changes both function content and binary-source correspondences by combining multiple source functions into one binary function. Jia et al.~\cite{jia2023inlining} and Abusabha et al.~\cite{abusabha2026inlining} investigate its effects on binary analysis. O2NMatcher~\cite{jia2022o2nmatcher} handles source-function combinations, and CI-Detector~\cite{jia2024crossinlining} addresses cross-inlining matching. CodeExtract~\cite{jia2024codeextract} uses code extraction and inlining to handle boundary differences, while BINALIGNER~\cite{zhu2026binaligner} aligns binaries across compilation environments. Building on contrastive learning~\cite{khosla2020supcon}, \tool{} accounts for multiple contributing source functions during training, with additional emphasis on major contributors.

\subsection{Structural Context and LLM-Assisted Matching}\label{sec-related-work-structure-llm}

Function context can help when the code of an individual function is insufficient to determine a match. Asteria-Pro~\cite{yang2024asteriapro} incorporates domain knowledge, while RevDecode~\cite{ren2025revdecode} and BinEnhance~\cite{wang2025binenhance} use dependencies and environmental relationships. BinStruct~\cite{zhang2025binstruct} and Modubin~\cite{yu2026modubin} recover file or module structure. Valenzuela and Kinder~\cite{valenzuela2026callgraphs} examine when call-graph context benefits different binary-analysis tasks. %
BinaryAI~\cite{jiang2024binaryai} uses link-time locality to obtain confident matches. \tool{}, however, advances the positional information not only to perform filtering but also to supplement candidates for unresolved functions.\space

LLMs have also been explored for function matching. Co$^{2}$FuLL~\cite{dong2025co2full} retrieves candidates using function content and context, and then verifies them with an LLM. Practical Source Code Recovery~\cite{gagnon2026source} retrieves source code using strings, constants, and call clues, and propagates high-confidence matches along call relations. \tool{} combines LLM-assisted matching with positional information to supplement candidates for unresolved binary functions beyond the initial retrieval results.

\section{Conclusion}\label{sec-conclusion}

We proposed \tool{}, a function-level binary-to-source matching method. %
\tool{} combines full-function embeddings with composition-aware training to retrieve source-function candidates, accounting for long functions and the multiple source compositions introduced by inlining. It then uses backend LLMs to select and review matches, while positional information helps filter inconsistent matches and supplement candidates for unresolved binary functions, extending the search beyond the initial retrieval candidates. Experiments %
showed that \tool{}, using different backend LLMs, can achieve higher overall precision, recall and F1 scores than the SOTA competitors. The proposed techniques, such as composition-aware training and candidate supplementation, are also well evaluated. \tool{} also exhibits acceptable efficiency and cost even for function matching on large software and TPLs.

\section*{Data Availability}\label{sec-data-availability}
The replication package for this work will be made publicly available
upon formal publication of the peer-reviewed version of this work.

\bibliographystyle{ACM-Reference-Format}
\bibliography{references}

\end{document}